\documentclass[preprint,12pt]{elsarticle}
\biboptions{square,comma,sort&compress}   

\usepackage{amssymb,amsmath,amsthm,epsfig,lineno}
\usepackage{hyperref}

\journal{Energy and Buildings}

\begin{document}

\begin{frontmatter}


\title{A feasibility study of passive ventilation via origami-driven stack effect}
\author[NYUAD]{Ahmad F.\ Zueter\corref{cor1}}
\author[KFUPMbio,KFUPMIRC]{Ahmad S.\ Dalaq}
\author[NYU]{Mohammed F.\ Daqaq}

\cortext[cor1]{Corresponding author: \href{}{ahmad.zueter@nyu.edu}}

\address[NYUAD]{Department of Mechanical Engineering, New York University Abu Dhabi, Abu Dhabi, United Arab Emirates}
\address[KFUPMbio]{Bioengineering Department, King Fahd University of Petroleum \& Minerals (KFUPM), Dhahran, Saudi Arabia}
\address[KFUPMIRC]{IRC Center for Biosystems and Machines, King Fahd University of Petroleum \& Minerals (KFUPM), Dhahran, Saudi Arabia}
\address[NYU]{Department of Mechanical Engineering, Tandon School of Engineering, New York University, Brooklyn, NY, United States of America}

\begin{abstract}
This study introduces an innovative ventilation system, which leverages an origami-inspired structure to improve and regulate natural airflow driven by the stack effect, with applications in underground mines and buildings. The proposed system retrofits chimneys and/or exhaust risers with expandable origami units that dynamically modulate their geometric features (height and vent area) to control buoyancy-driven ventilation independent of external wind or solar conditions.
The efficacy of the proposed concept is evaluated using a three-dimensional computational model that focuses on how the geometric design of the deployable stack and varying atmospheric conditions affect the volumetric airflow through the stack in both its fully expanded and contracted states.
The results show that the ventilation rate increases by up to 25\% with each doubling of the height of the origami stack. %
These findings highlight the potential of leveraging origami-inspired structures for adaptable and controllable energy-efficient ventilation, which is particularly beneficial in energy-intensive applications and/or remote, off-grid locations.

\end{abstract}

\begin{graphicalabstract}
\centering
\includegraphics{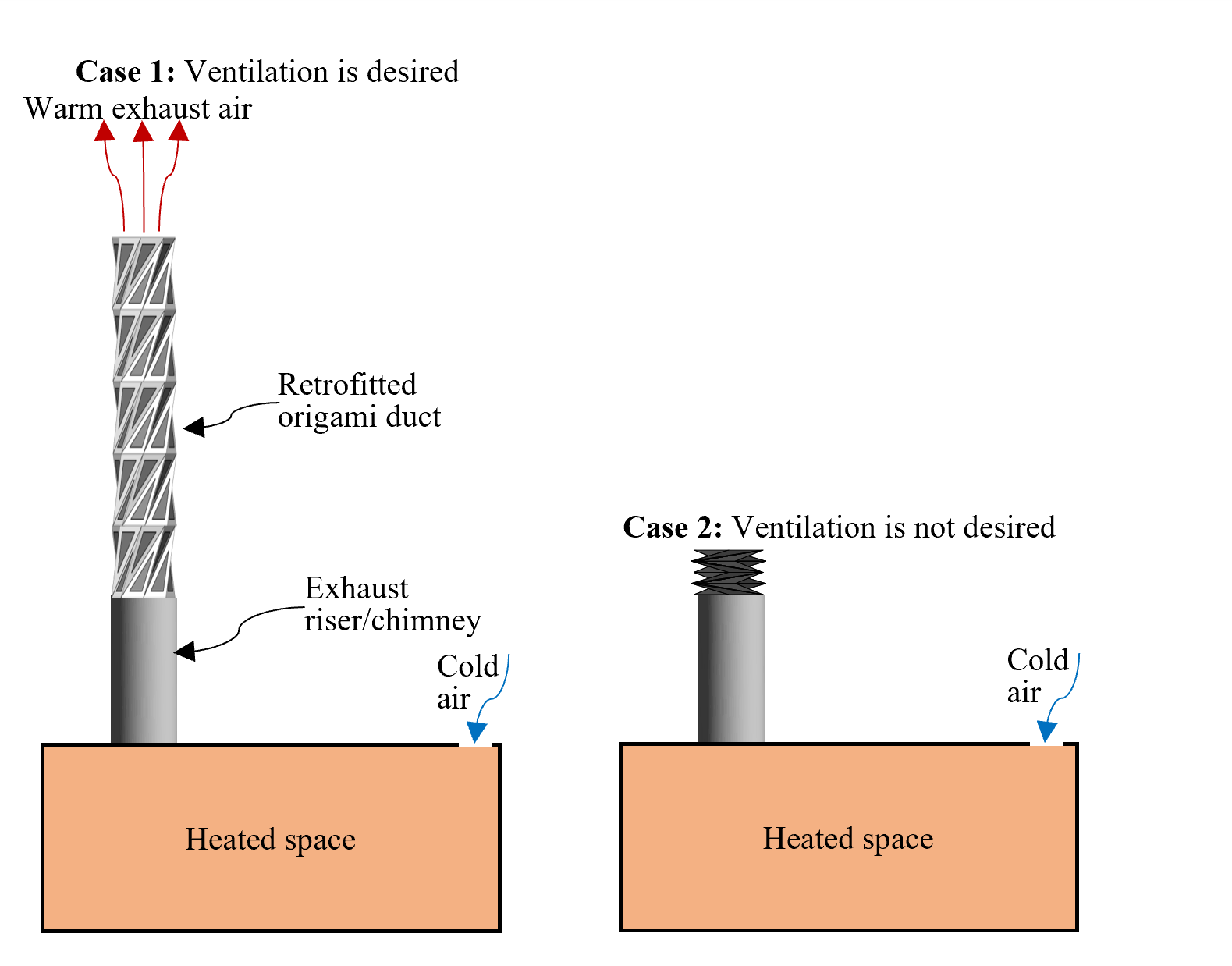}\\
\end{graphicalabstract}

\begin{highlights}
\item A novel ventilation system using origami structures to regulate natural airflow.
\item Based on chimneys retrofitted with origami units that dynamically modulate their geometry.
\item Achieves a minimum of threefold ventilation control by tuning the origami design parameters.
\item Ventilation increases by approximately 25\% with each doubling of the origami stack height.
\item Significant pressure losses inside origami structures with low number of triangular panels
\end{highlights}


\end{frontmatter}


\section{Introduction}\label{sec:intro}
Ventilation accounts for a substantial share of energy consumption in multiple sectors. In underground mining operations, for instance, ventilation can account for up to 50\% of total operational energy costs~\cite{Demirel2018}, primarily due to the continuous need for fresh air supply and heat removal. 
Similarly, in the building sector, heating, ventilation, and air conditioning (HVAC) systems contribute to nearly 40\% of global building energy use~\cite{InternationalEnergyAgency2022}. 
These energy demands are especially challenging in off-grid areas, where access to reliable electricity is limited and costly. As a result, there is a growing need for passive, sustainable indoor ventilation solutions that do not depend on external power sources. 

The stack effect, also known as the chimney effect, offers one solution for natural and sustainable ventilation. It is driven by buoyancy forces that arise from the temperature-induced difference in density between indoor and outdoor air. In cold regions, for example, the warmer and less dense indoor air rises, creating a vertical pressure gradient that draws cooler air in through lower openings and expels warm air through upper ones.  The volumetric flow rate of air, $\dot{V}$, due to the stack effect is directly proportional to the height of the stack, $H$~\cite{Song2014}, the difference in indoor-outdoor temperature, $\Delta T$~\cite{Yang2023}, and the vent's cross-sectional flow area,~$A$~\cite{Yin2024}. 

Numerous studies have explored the stack effect as a natural ventilation mechanism~\cite{Wong2004,Min2009}. For example, Hughes and AbdulGhani~\cite{Hughes2008} demonstrated that, under appropriate conditions, stack-effect-driven ventilation meets British standards. %
However, a persistent challenge remains to determine how to effectively control this passive strategy to consistently ensure comfort and safety across various climates and building designs. 

Toward this end, several passive strategies have been developed to improve the efficacy of natural ventilation mechanisms. These include the use of solar walls in cold climates~\cite{Khanal2011,Ghamari2024} and windcatching vents in hot ones~\cite{Chohan2022,Filis2021}. 
Solar walls, or Trombe walls, utilize solar heating of a narrow air cavity to increase the temperature differential, $\Delta T$, thus improving the ventilation rate in cold climates~\cite{Yu2021,Sanij2015,Khrystyna2024}. Windcatchers, on the other hand, are oriented to catch the prevailing wind and pass it through cooler areas like basements before channeling it to living areas. They have been shown to improve ventilation rates by up to 15\%~\cite{Obeidat2023}, while significantly lowering cooling loads~\cite{Nejat2021,Nejat2024}. Hybrid systems that combine solar walls with windcatchers or other active, energy-intensive components have also been proposed to further optimize indoor comfort~\cite{Saifi2024,Tumer2013,Bosu2023}.

 %
 %
Since all previously proposed passive ventilation methods rely on fixed architectural features~\cite{Laurini2018}, specifically stack height, $H$, and vent area, $A$, their performance can only be adjusted by varying the temperature difference, $\Delta T$. As a result, the effectiveness of these strategies is heavily dependent on environmental factors such as sunlight and wind, which can be intermittent or absent. Therefore, there is a need for passive ventilation solutions that offer an additional level of control independent of environmental conditions. One promising approach to address this limitation is to develop stacks with adjustable architecture, eliminating the constraints of fixed stack height, $H$, and vent area, $A$.

\begin{figure}[h!]
\centering
\includegraphics{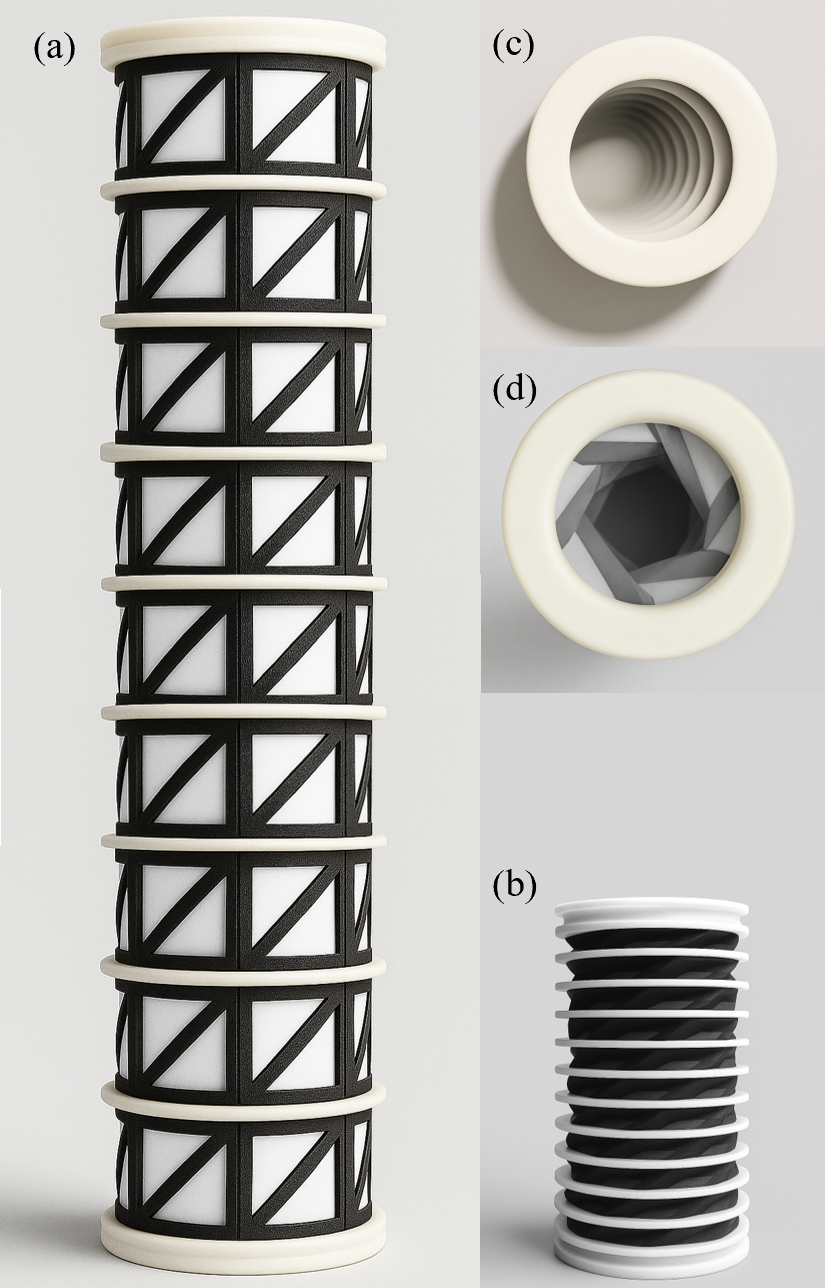}
\caption{Kresling origami in (a) expanded state and (b) contracted state, with the corresponding top view of the (c) expanded state and (d) contracted state.}\label{fig:origami_physical}
\end{figure}

To achieve this goal, we propose to rely on a new field of mechanics that concerns the design of adaptable and collapsible structures by incorporating origami principles \cite{Masana2024}. In particular, an origami pattern, known as the Kresling pattern \cite{Dalaq2022} has been used to build collapsible columns (stacks) whose height and cross-sectional area can be controlled by applying an external force. Fig.~\ref{fig:origami_physical} depicts one such stack in its fully expanded and contracted states, demonstrating that significant changes in its height (compare Fig.~\ref{fig:origami_physical} (a) and (b)) and cross-sectional area ((compare Fig.~\ref{fig:origami_physical} (c) and (d)) can be realized with minimal energy input \cite{Khazaaleh2024b,Khazaaleh2024}. These changes can be used to control the volumetric flow rate of air, $\dot{V}$, and therewith the ventilation rates for a given fixed temperature gradients, $\Delta T$.

The potential of using origami-based structures for airflow management has been discussed in various domains~\cite{Akhtar2023}. 
In healthcare, Zhang et al.~\cite{Zhang2024} and Kim et al.~\cite{Kim2021} developed origami-inspired portable ventilators, while Heatherington et al.~\cite{Heatherington2022} introduced origami nasal covers to mitigate aerosol transmission. 
Pesenti et al.~\cite{Pesenti2018} demonstrated the potential of deployable Kresling structures for thermal management in indoor environments. %
Zhang et al.~\cite{Zhang2025_shading} integrated photothermal-responsive Kresling units for solar modulation and ventilation control. %
Other innovations include origami-inspired window systems for dynamic airflow and sound insulation were proposed by Jin et al.~\cite{Jin2023}.  %
More recently, Zhang et al.~\cite{Zhang2025_drag} investigated the aerodynamic drag characteristics of Kresling designs to assess their feasibility in ventilation applications. %

\begin{figure}[h!]
\centering
\includegraphics[width=\textwidth]{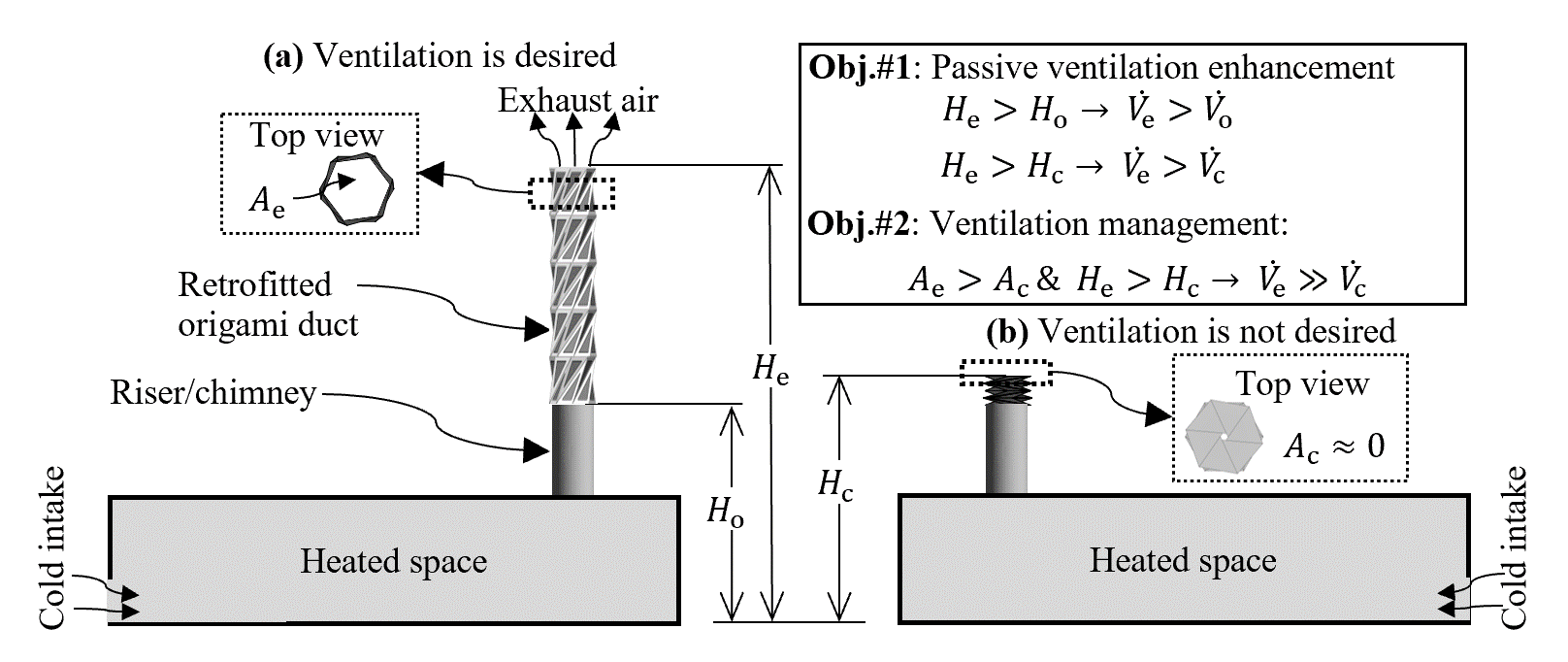}
\caption{Proposed concept for enhancing and managing passive ventilation through (a) expanding and (b) contracting a retrofitted origami duct. Subscripts `e' and `c' indicate expanded and contracted states of origami, respectively, with subscript `o' referring to original system prior to retrofitting origami.}\label{fig:concept}
\end{figure}

In this study, we present a novel passive ventilation enhancement system that increases airflow by passively modulating the effective stack height through a Kresling deployable mechanism retrofitted onto chimneys or exhaust ducts, as illustrated in Fig.~\ref{fig:concept}. 
Unlike conventional devices such as valves, which only regulate or restrict flow, this design enhances ventilation rates beyond existing limits by effectively extending the stack height, independent of external wind or solar conditions. 
As the first investigation of its kind, this work is limited to computational analysis, focusing on how the geometric configuration of the deployable stack and varying atmospheric conditions influence the volumetric airflow in both fully expanded and contracted states. 
The primary objective is to evaluate the feasibility of the concept and to identify the key design parameters that govern the achievable airflow enhancement.

The remainder of this paper is organized as follows: Section~\ref{sec:conceptual_methodology} presents the conceptual methodology.
and delves into the design of Kresling origami cells. 
Section~\ref{sec:analytical} formulates the underlying physics governing the stack effect phenomenon.
Section~\ref{sec:mathematical_methodology} describes the mathematical formulation, including computational fluid dynamics (CFD) modeling and analytical analysis of buoyancy-driven airflow. %
Section~\ref{sec:results} presents the results and evaluates the system's effectiveness. %
Finally, conclusions are drawn in Section~\ref{sec:conclusion}. %

\section{Conceptualization}\label{sec:conceptual_methodology}
\subsection{Overview}
This study introduces a novel concept for utilizing the stack effect in passive ventilation through architectural modifications inspired by origami structures. Specifically, our approach involves retrofitting existing exhaust ducts or chimneys with adaptable Kresling origami units capable of simultaneously altering both the stack height, $H$, and the outlet vent area, $A$, as illustrated in Fig.~\ref{fig:concept}. %
These structures are actuated vertically via an external force or torque applied through a motor mounted at the base of the stack. %
Upon actuation, the geometry of individual Kresling cells changes, resulting in adjustments to both the overall stack height and the vent area. In particular, the origami stack can be designed such that its height, $H_\text{e}$, and the outflow cross-sectional area, $A_\text{e}$, in the expanded position are substantially greater than its height, $H_\text{c}$, and the outflow cross-sectional area, $A_\text{c}$, in the contracted position as shown with a physical origami stack in Fig.~\ref{fig:origami_physical} and illustrated in Fig.~\ref{fig:concept}. %
This geometric transformation allows for meaningful modulation of the ventilation rate, offering a high degree of passive tunability with minimal mechanical complexity. %

\subsection{The Origami Stack}~\label{sec:origami_design}
\begin{figure}[h!]
\centering
\includegraphics[width=\textwidth]{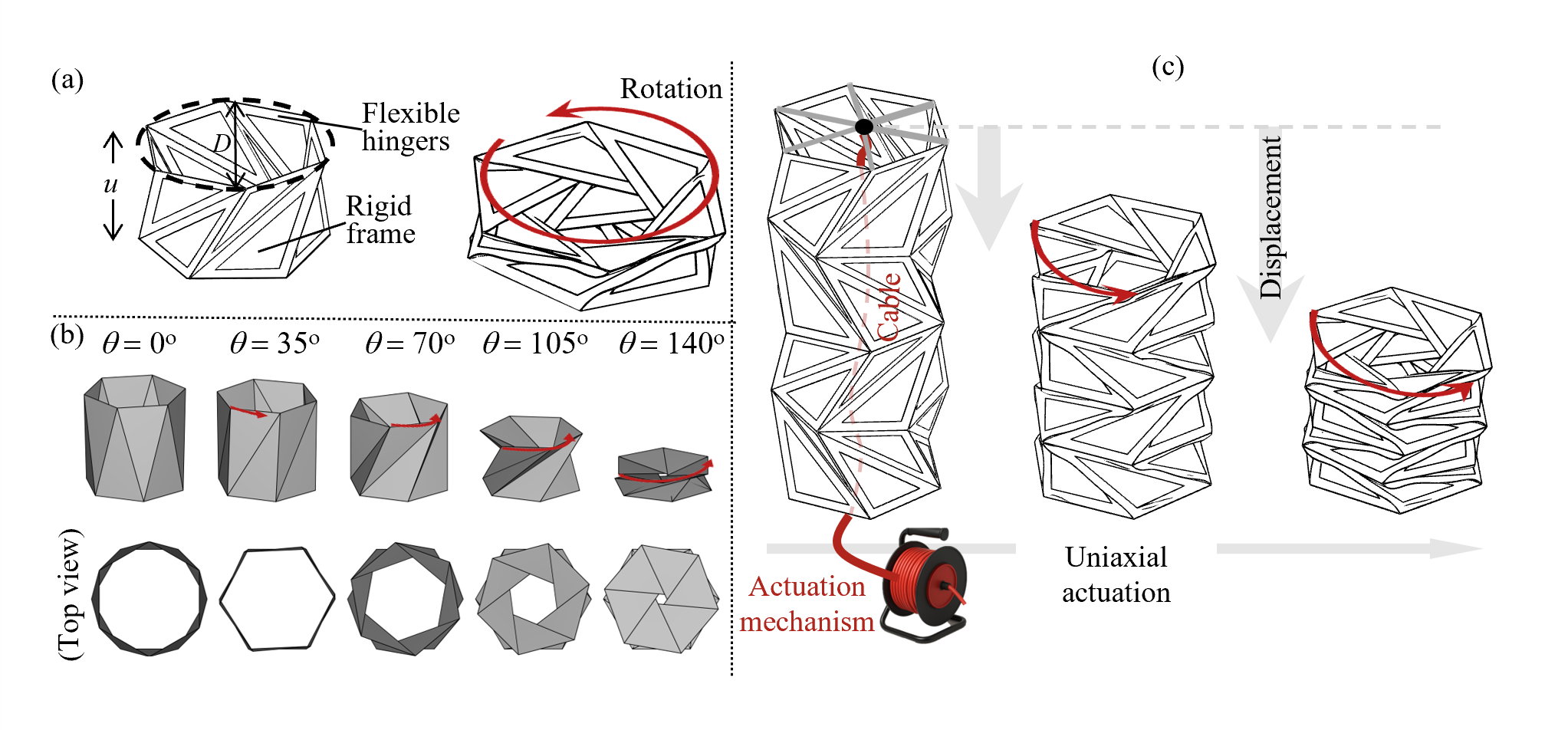}
\caption{Illustration of actuation mechanism in our proposed concept: (a) Components of a single origami unit cell, (b) compression of a unit upon rotation, and (c) compression of a stack of origami units}\label{fig:actuation}
\end{figure}

The origami stack is formed by connecting multiple Kresling origami unit cells in series. Each unit cell is a cylindrical structure composed of a specific number of identical triangular panels arranged in a zigzag pattern as shown in Fig.~\ref{fig:actuation}(a).
Each triangle is highly rigid and connected to adjacent triangles using flexible materials or hinges (see ref~\cite{Masana2024} for possible ways to fabricate such unit cells).  When one side of a cylinder is fixed while the other side is subjected to a force or torque, the unit cell folds as shown in Fig.~\ref{fig:actuation}(b) resulting in variations in the cell's height and cross-sectional area. Similarly, when the bottom end of the stack is fixed to the exhaust duct or chimney, while the upper end is actuated in the vertical direction using a cable-motor system as shown in Fig.~\ref{fig:actuation}(c), the force exerted by the cable causes the unit cells to fold and the stack to compress, resulting in changes in the effective height and cross-sectional area of the stack.

\subsection{Cell Design Parameters and Operation Limits}
\begin{figure}[h!]
\centering
\includegraphics{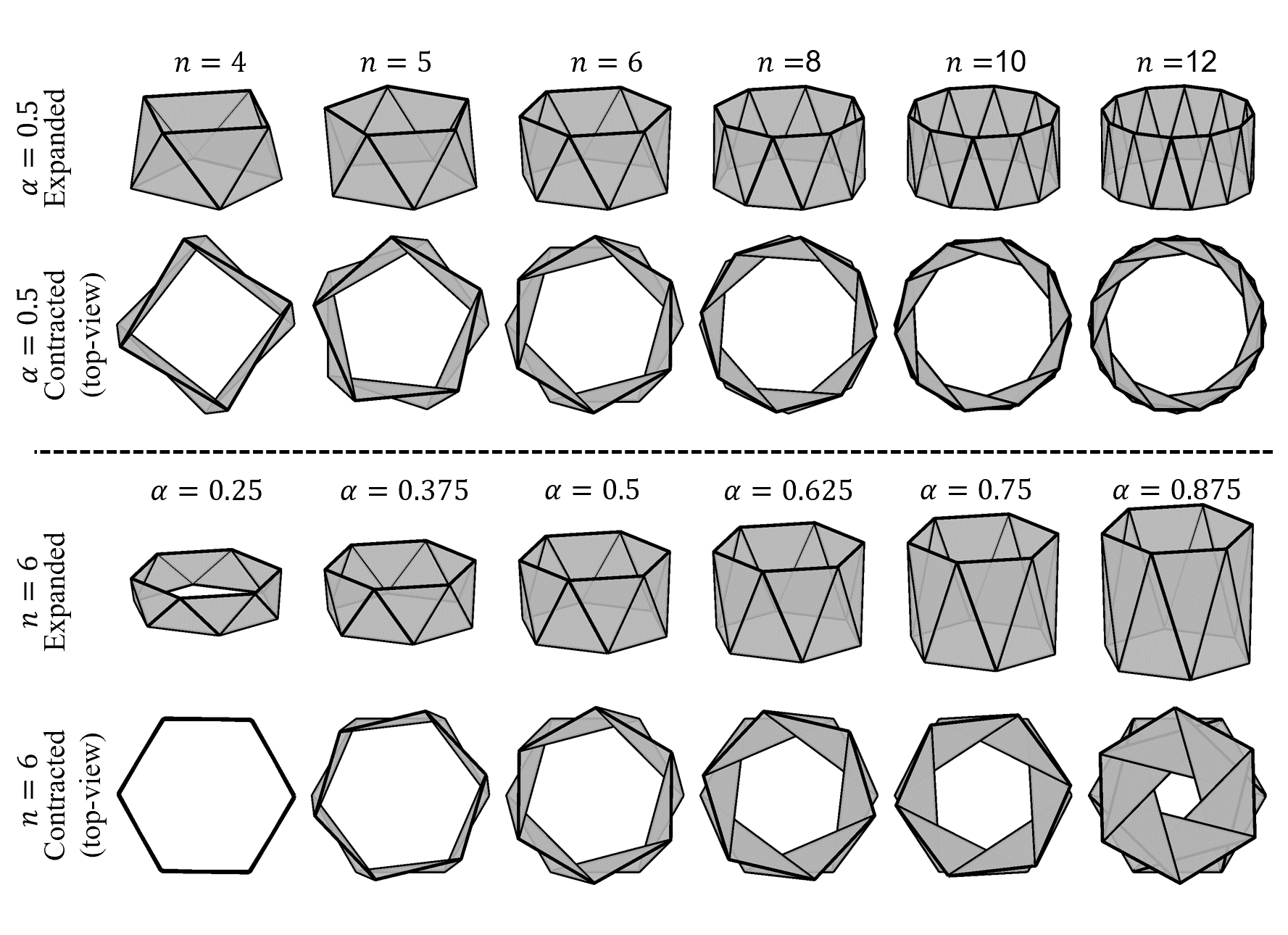}
\caption{Impact of design parameters $n$ (number of polygon sides at the top/bottom faces) and $\alpha=u/D$ (cell aspect ratio)}\label{fig:geometry_parameters}
\end{figure}

As shown in Fig. \ref{fig:actuation}(a), each unit cell that forms the stack is characterized by two geometric design parameters. These are $i)$ the number of sides, $n$, of the polygon forming the cross section of the unit cell, and, $ii)$  the aspect ratio $\alpha=u/D$ of the cell, which represents the ratio between the pre-deformation height, $u$, and the diameter, $D$, of the cylindrical envelope that bounds the cross-section. These parameters not only influence the shape and height of the final stack, but also its cross-sectional area and folding behavior.  In what follows, we discuss the influence of these parameters on the geometry and folding behavior of the unit cell:
\begin{itemize}
\item Effect of $n$: As shown in Fig.~\ref{fig:geometry_parameters}, for a given $\alpha$, the area of the cross section increases in both the expanded and contracted positions when $n$ increases. The cross-sectional area approaches that of a circle when $n$ approaches infinity.
\item Effect of $\alpha$: As shown in Fig.~\ref{fig:geometry_parameters}, for a given $n$, the cylinder becomes more slender as $\alpha$ increases. Thus, less number of unit cells become necessary to form stacks of similar height. Furthermore, the range of rotation to full contraction increases as $\alpha$ increases therewith reducing the cross-sectional area in the contracted position.
\end{itemize}

Each unit cell can undergo rotation to a maximum theoretical limit, $\theta_{\text{max}}$. 
This limit is governed by the smaller value of two angles: $i)$ $\theta_n$ which defines the crumpling limit and depends on $n$, and $ii)$ $\theta_\alpha$ which denotes the angle at which the cell flattens and its vertical height reduces to zero. Assuming negligible panel thickness, the maximum possible rotation can be mathematically expressed as:
\begin{equation}\label{eq:theta_rotation}
\theta_{\text{max}} = \min\Big(
\underbrace{\pi\frac{n-1}{n}}_{\substack{\theta_n}},~
\underbrace{2 \sin^{-1}(\alpha)}_{\substack{\theta_\alpha}}
\Big).
\end{equation}

In cells with near-zero aspect ratios, the maximum rotation angle is primarily governed by $\theta_\alpha$. As the aspect ratio approaches unity, however, $\theta_{\text{max}}$ becomes limited by panel crumpling rather than geometric folding. %
Ideally, the maximum possible rotation during folding is 180$^\circ$ achieved as $n$ approaches infinity and the cell aspect ratio approaches unity.

\section{Underlying Physics of the Stack Effect}\label{sec:analytical}
\begin{figure}[h!]
\includegraphics[width=\textwidth]{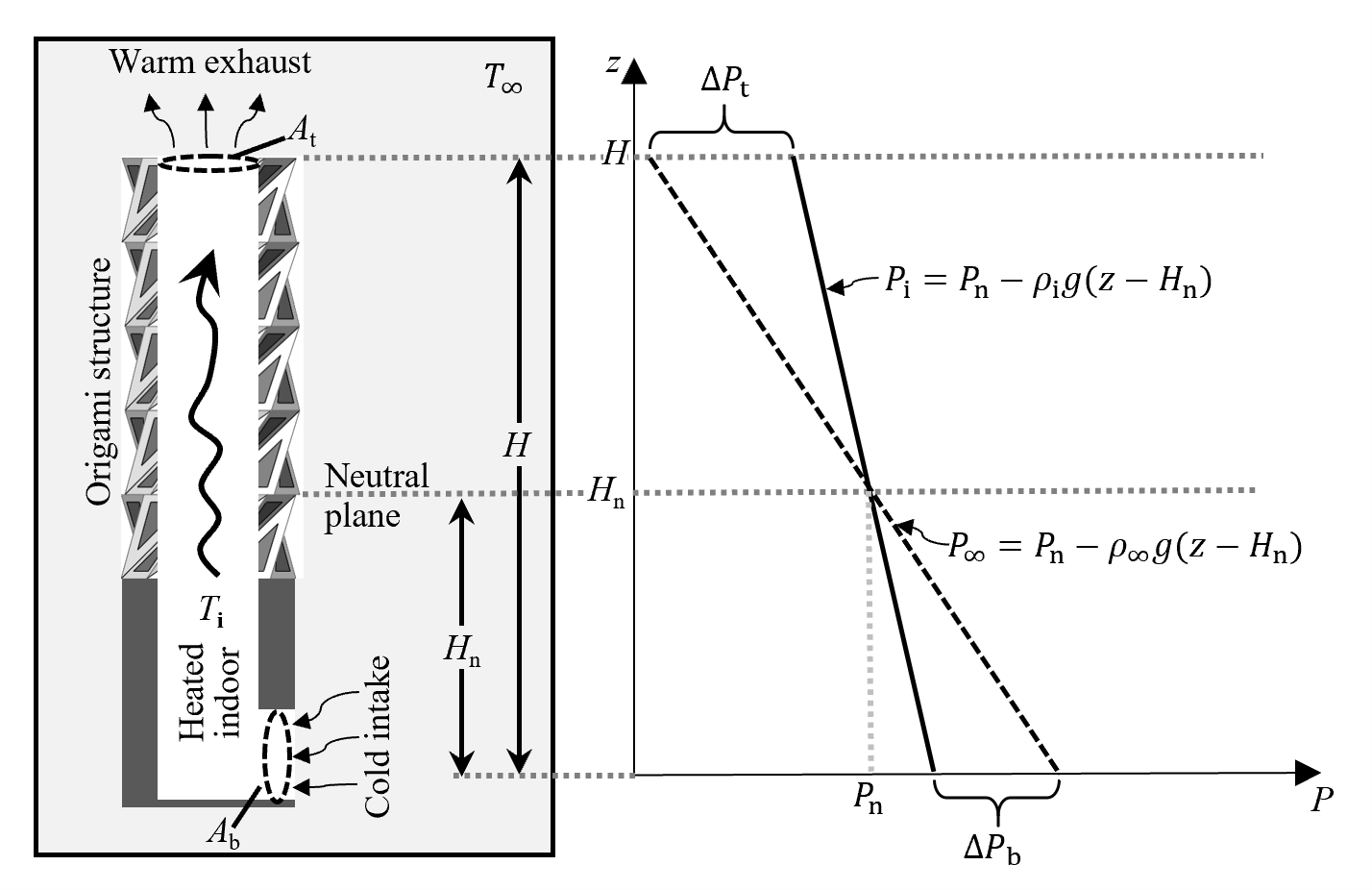}
\caption{Illustration of the underlying physics of the stack effect phenomena in cold climates}\label{fig:analytical}
\end{figure}

The stack effect is caused by temperature-induced variations in air density between an indoor space (denoted here with subscript `i') and an outdoor space (denoted with subscript `$\infty$'). 
This variation in air density creates a different hydrostatic pressure profile in indoor and outdoor spaces, as shown in Fig.~\ref{fig:analytical}, promoting air circulation across the indoor space. In this section, we present a simple analytical model to understand the stack effect and predict the mass flow entering an arbitrary tower with varying chimney height (i.e., the origami retrofit). We focus on the forward stack effect with a heated indoor space located in cold climates, as shown in Fig.~\ref{fig:analytical} (details on the reverse stack effect can be found in~\cite{klote1992design,klote_general_1991,mijorski_stack_2016}). 

The model invokes the following assumptions:
\begin{itemize}
\item[$\bullet$] Steady-state conditions are assumed.

\item[$\bullet$] Indoor and outdoor air are each uniform in temperature (\(T_\text{i}\) and \(T_\infty\)), and their densities \( \rho_\text{i} \) and \( \rho_\infty \) are calculated using ideal gas equation as $\rho = P_\text{atmospheric}/(R T)$, where $R$ is the specific ideal gas constant for air.

\item There is a negligible temperature gradient between the indoor and outdoor environments; the temperature transitions abruptly from one state to the other.

\item[$\bullet$] The only driving force is buoyancy. No convective or mechanical pressurization (e.g., wind, fans, or compressors) is considered.
\end{itemize}

Assuming these conditions, we consider two openings located at the top and bottom of a heated building, denoted by subscripts t (top) and b (bottom), and separated by a vertical distance $H$, known as the \textit{stack height}. Since cold outside air is denser than the warm indoor air, the pressure at the bottom opening is typically higher outside than inside. Furthermore, because pressure varies linearly with height, $z$, for both the indoor and outdoor air, and since $\rho_\infty > \rho_\text{i}$, the pressure profiles intersect at some height $z = H_\text{n}$, where both the indoor and outdoor have equal pressure values of $P_\text{n}$. Above this neutral height, the outdoor pressure exceeds the indoor pressure. 

Using this understanding, energy conservation at the bottom opening ($z=0$) can be expressed as a balance between the pressure difference arising from the hydrostatic head (see Fig.~\ref{fig:analytical}) with air acceleration towards the opening and pressure losses attributed to entrance effects, $\Delta P_\text{e}$, as  
\begin{equation}\label{eq:pressure_bottom}
\Delta P_\text{b} = \underbrace{g H_\text{n} (\rho_\infty - \rho_\text{i})}_{\text{hydrostatic pressure}}~~~ = \underbrace{\frac{1}{2}\rho_\infty \overline{v}_\text{b}^2}_{\text{dynamic pressure}} + ~~~\underbrace{\Delta P_\text{e}}_{\text{losses}},
\end{equation}
where $\Delta P_\text{e} =\frac{1}{2} K_\text{e}\rho_\infty v_\text{b}^2$,
with $K_\text{e}$ being the entrance loss coefficient, dependent on the opening geometry~\cite{cengel_fluid_2014}.

Similarly, energy balance at the top opening ($z=H$) can be expressed as
\begin{equation}\label{eq:pressure_top}
\Delta P_\text{t} = \underbrace{ g (H-H_\text{n})(\rho_\infty - \rho_\text{i})}_{\text{hydrostatic pressure}}~~~= \underbrace{\frac{1}{2}\rho_\text{i}\overline{v}^2_\text{t}}_{\text{dynamic pressure}} + ~~~\underbrace{\Delta P_\text{f}}_{\text{losses}},
\end{equation}
where $\Delta P_\text{f}$ represents friction pressure losses along the exhaust duct height, $H_\text{d}$, as~\cite{cengel_fluid_2014}
\begin{equation}\label{eq:pressure_friction}
    \Delta P_\text{f} = f \frac{H_\text{d}}{D_\text{H}}\frac{\rho_\text{i}v_\text{t}^2}{2}
\end{equation}
where $f$ and $D_\text{H}$ represent the flow friction factor and hydraulic diameter.

The velocities $\overline{v}_\text{t}$ and $\overline{v}_\text{b}$ can be correlated with each other by applying mass conservation as 
\begin{equation}\label{eq:mass_conservation}
    \dot{m} = \overline{v}_\text{b}\rho_\infty A_\text{b} = \overline{v}_\text{t}\rho_\text{i} A_\text{t},
\end{equation}
where $A_\text{t}$ and  $A_\text{b}$ refer to the cross-sectional areas of the top and bottom openings. 

Combining Eq.~\eqref{eq:pressure_bottom},~\eqref{eq:pressure_top}, and~\eqref{eq:mass_conservation}, we obtain the following expression for $H_\text{n}$: 

\begin{equation}\label{eq:neutral_height}
H_\text{n} = H_\text{n,ideal} + \frac{\hat{A}^2 \hat{T} \Delta P_\text{e}- \Delta P_\text{f}}{g (\rho_\infty - \rho_\text{i})(\hat{A}^2 \hat{T} + 1)}
\end{equation}
where $\hat{A}=A_\text{b}/A_\text{t}$ and $\hat{T}= T_\text{i}/T_\infty$. The ideal neutral height ($H_\text{n,ideal}$) represents the neutral height position without losses, given as 
\begin{equation}
        H_\text{n,ideal} =  \frac{H}{1+\hat{T} \hat{A}^2}.
\end{equation}
Note that the sign of $\Delta P_\text{e}$ in Eq.~\eqref{eq:neutral_height} is positive, indicating that the neutral height must rise in order to increase $\Delta P_\text{b}$ and thereby compensate for the entrance losses, as shown in Eq.~\eqref{eq:pressure_bottom}.
In contrast, the sign of $\Delta P_\text{f}$ in Eq.~\eqref{eq:neutral_height} is negative, indicating that the neutral height must decrease to increase $\Delta P_\text{t}$ and thus offset the frictional losses, as shown in Eq.~\eqref{eq:pressure_top}. 

Once $H_\text{n}$, $v_\text{b}$, and $v_\text{t}$ are determined from Eq.~\eqref{eq:pressure_bottom}, Eq.~\eqref{eq:pressure_top}, and Eq.~\eqref{eq:mass_conservation}, the overall ventilation rate is then evaluated by applying mass conservation through either opening, as expressed in Eq.~\eqref{eq:mass_conservation}.

\section{Evaluation Methodology}\label{sec:mathematical_methodology}
The proposed passive ventilation concept is evaluated using a three-dimensional (3D) computational model that captures key features of the stack effect; namely the velocity profile and the ventilation flow rate.  The model, which incorporates the complex geometry of the origami structures and the associated pressure losses, is validated for simplified cases using experimental data available in the literature. 

\begin{figure}[t!]
\centering
\includegraphics{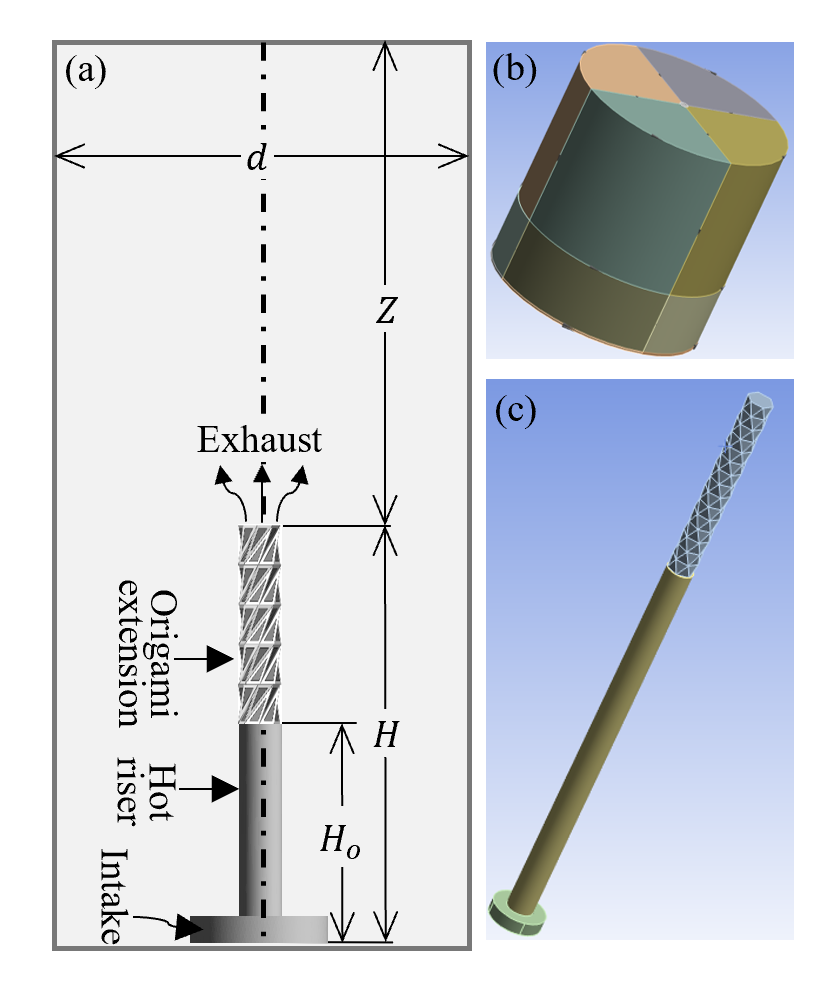}
\caption{Computational modeling of the proposed concept: (a) Not-to-scale schematic of the computational domain, (b) entire computational domain for CFD analysis, and (c) origami and riser walls in the model.}\label{fig:methodology}
\end{figure}

\begin{figure}[h!]
\centering
\includegraphics[width=13.9cm]{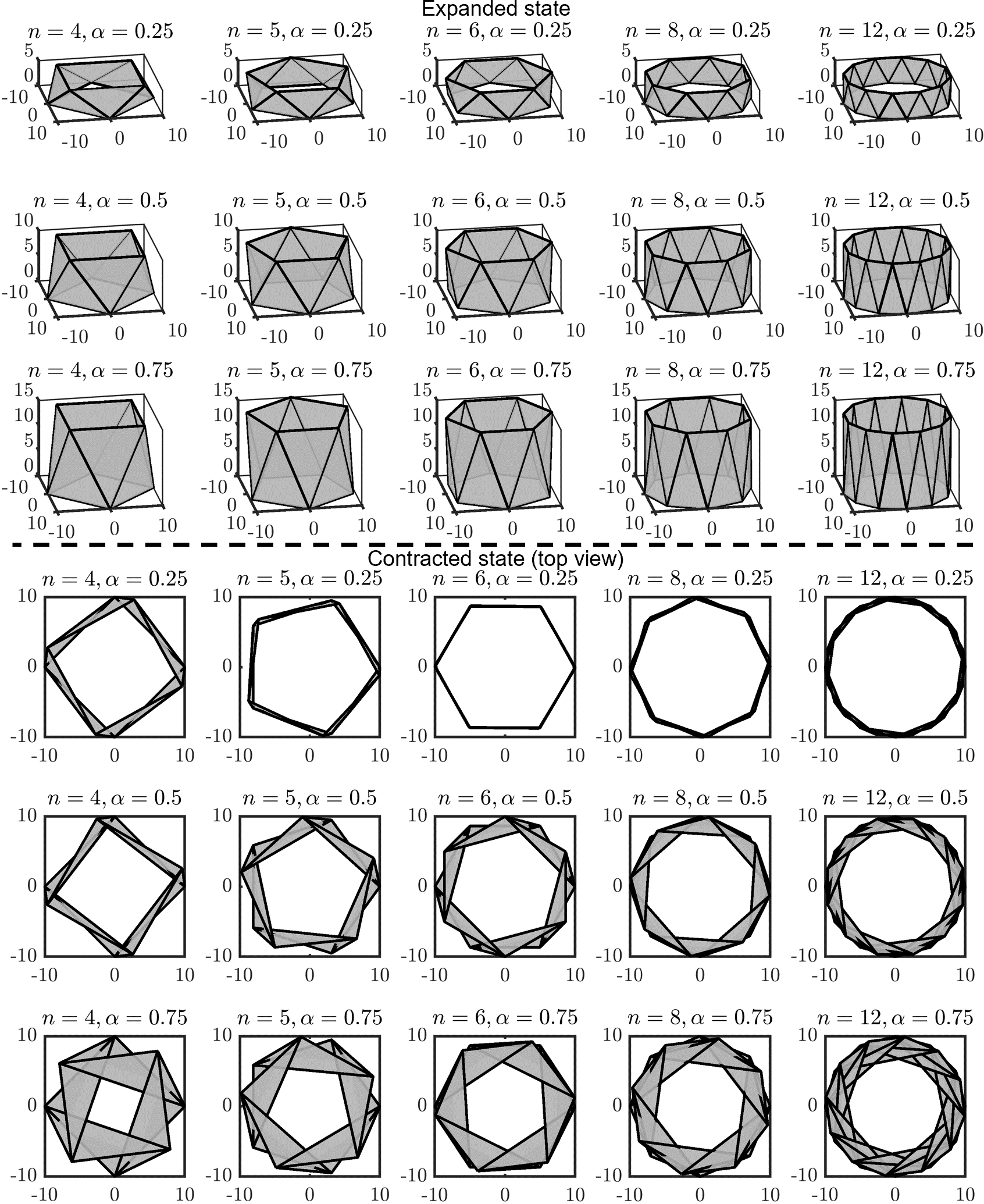}
\caption{Selected cases of unit cells used to construct the Kresling origami extension - axes units in centimeters.}\label{fig:appendix_origami_Cases}
\end{figure}

The computational domain shown in Fig. \ref{fig:methodology} takes the shape of a cylinder of diameter, $d$, and height, $Z+H$. It contains a heated riser (chimney) with a fixed diameter $D = 20$~cm, and a height of $H_\mathrm{o} = 310$~cm. The riser temperature is fixed at 20$^\circ$C to induce natural convection, drawing air from the colder ambient environment whose temperature is fixed at $T_{\infty}$ ($T_{\infty}$ is varied depending on the simulation). An origami extension, composed of stacked origami cells, is mounted atop the riser extending the total riser height to $H$ in the expanded position of the origami stack. 
The number of stacked cells, $N$, is set to 5, 10, and 15 for $\alpha = 0.25$, $0.5$, and $0.75$, respectively, such that the total stack height increases by an equal value of 150cm in all cases.
Several unit cell geometries with various values of $n$ are considered. These cases are shown in Fig.~\ref{fig:appendix_origami_Cases} in the expanded and contracted states. 

\subsection{Computational model}\label{sec:fluid_cfd_methodology}
 A 3D transient computational model is constructed based on the conservation equations of mass, momentum, and energy. Air is modeled as an ideal gas to fully couple pressure and temperature with density variations. 
No-slip and adiabatic boundary conditions are applied at all walls, allowing the analysis to isolate the impact of the origami extension without introducing additional interfering parameters from heat transfer through the solid boundaries, which is beyond the scope of this study. %
Moreover, the Shear Stress Transport (SST) $k$--$\omega$ turbulence model is employed, as the Reynolds number of the flow inside the riser and origami extension exceeds $1 \times 10^5$. %
The full formulation of the Reynolds-Averaged Navier–Stokes (RANS) equations and the applied boundary conditions are detailed in \ref{Appendix:computational}. %

Various numerical parameters and solution strategies are adopted to ensure both stability and accuracy of the simulations. 
A structured mesh is generated throughout the computational domain, using hexahedral elements outside the ventilation system and refined prism layers within the origami layers to accurately resolve the complex geometry of the origami structure. 
Given the strong coupling between the velocity and pressure fields, the stability of the solution was maintained by solving these variables implicitly and simultaneously. 
Under-relaxation factors were applied to the velocity, pressure, and turbulence fields to further enhance numerical stability. 
In addition, second-order accurate schemes were employed for the discretization of momentum, temperature, and other flow variables. 
Mesh generation and numerical computations are carried out using ANSYS FLUENT (R2023), as shown in Fig.~\ref{fig:methodology}(b-c). 

\begin{figure}[h!]
\centering
\includegraphics[width=\textwidth]{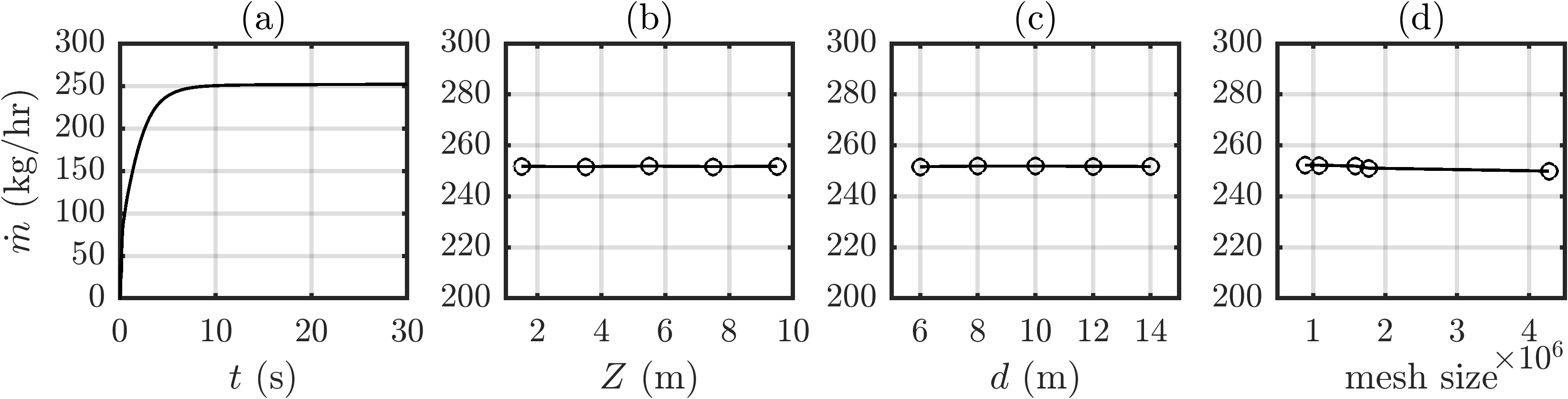}
\caption{a) Transient solution of ventilation rate until reaching steady state, b) independence analysis from domain height, c) independence analysis from domain diameter, and d) mesh independence.}\label{fig:independence}
\end{figure}

To ensure that physical results are not influenced by numerical parameters, a set of independence studies was carried out on a case with $N=15$ expanded origami units each with $n = 6$ and $\alpha = 0.5$, reaching a total stack height of $H = 4.5$ m, as shown in Fig.~\ref{fig:methodology}(b). %
The results of the transient simulation shown in Fig.~\ref{fig:independence}(a) indicate that the ventilation mass flow rate stabilizes and reaches a steady state after approximately 15 seconds. The subsequent parametric analysis shown in Fig.~\ref{fig:independence}(b,c) confirms that both the domain height above the origami extension, $Z$, and the domain diameter, $d$, are sufficiently large, as the outlet mass flow rate remains essentially constant across the tested values. %
While larger domains may be necessary if steady state is achieved over longer timescales, the short transient period observed here indicates that the chosen domain size does not influence the results. 
Furthermore, the mesh independence analysis shown in Fig.~\ref{fig:independence}(d) demonstrates negligible variation in the results across different mesh sizes, confirming that the numerical solution is not affected by further mesh refinement. %

Based on these results, for all subsequent simulations, the height of the domain measured from the top of the exhaust is fixed at $Z = 5.5$\,m, the diameter at $d = 10$\,m, and the mesh size contains between 1-2 million elements depending on the simulation case. Each case is simulated for 30 seconds to ensure that the response has reached its steady-state value. %

\subsection{Model validation}\label{sec:methodology_val}
\begin{figure}[t!]
\centering
\includegraphics[width=\textwidth]{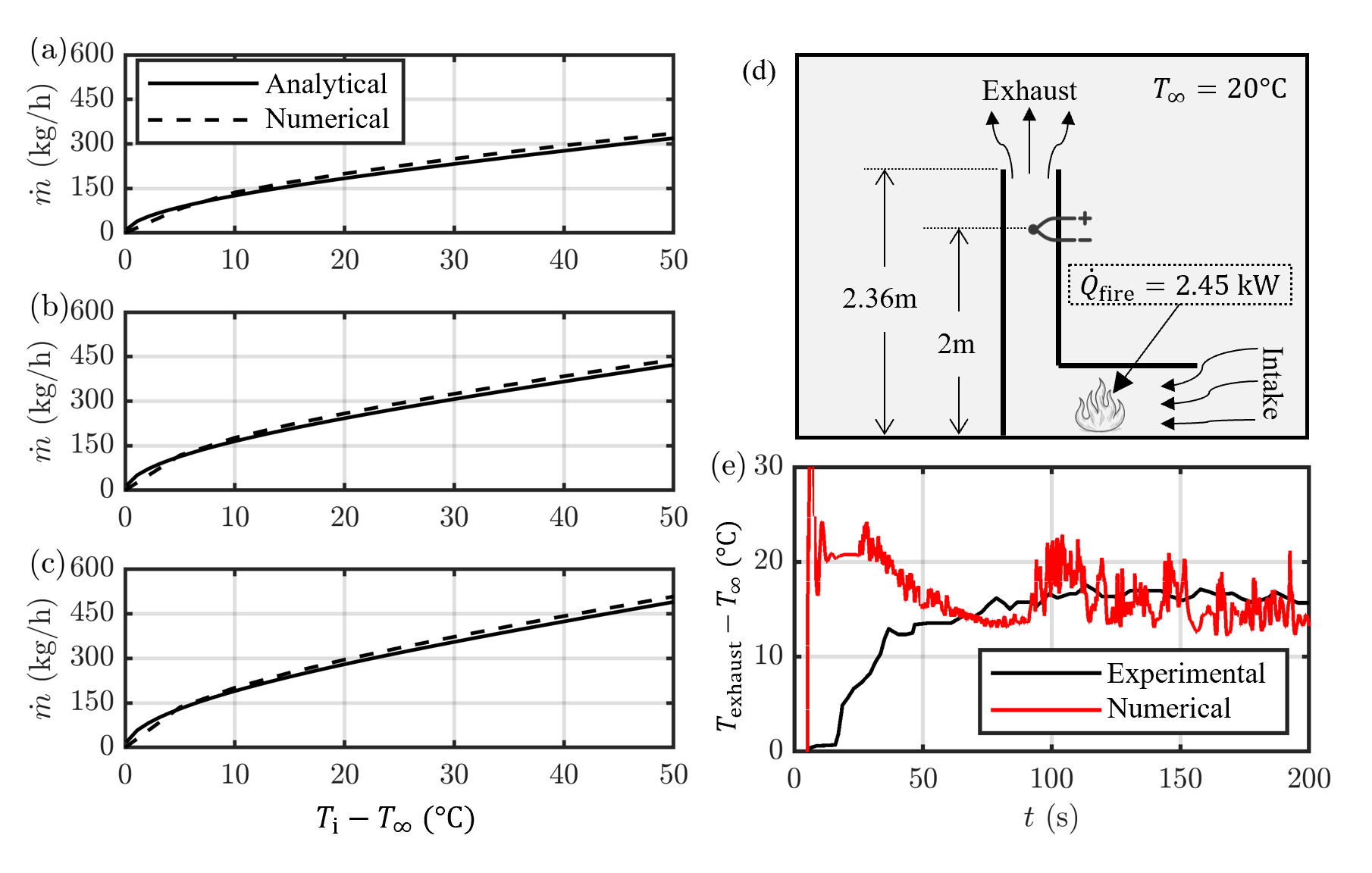}
\caption{Verification of the computational results with analytical solutions at (a) $H^*=1$, (b) $H^*=2$, and (c) $H^*=3$, followed by (d) a non-to-scale schematic of the experimental design of Zhang et al.~\cite{Zhang2006} and (e) computational results validation with the experimental data of Zhang et al.~\cite{Zhang2006}.}\label{fig:validation}
\end{figure}

To assess the accuracy of the computational model, its output is first compared with the analytical results, and then validated using a large-scale experiment from the literature, as shown in Fig.~\ref{fig:validation}. In both cases, the computational model  excludes the origami structure to match the assumptions of the analytical model and the conditions of the experimental setup. In other words, the computational domain is adapted to enable a fair comparison with both analytical and experimental benchmarks. %

The results of the computational model are first compared with the analytical solution for different stack heights, $H^* = H/H_\text{o}$ (with $H_\text{o} = 3$m) and indoor-outdoor temperature differences, $\Delta T=T_{\text{i}}- T_{\infty}$. The goal of these comparisons is to assess whether the computational model qualitatively reflects the analytical trends.
To this end, a total of 30 CFD simulations were conducted at three stack heights, namely $H^* = 1, 2, 3$, and ten temperature differentials incremented uniformly from $\Delta T = 5$ to $\Delta T = 50$, as shown in Fig.~\ref{fig:validation}(a-c). %
Both models show that the ventilation rate increases with $H^*$ and $\Delta T$. 
However, while both models show good agreement at low stack heights, the analytical model overpredicts the ventilation rate obtained using the CFD model as $H^*$ increases. This overprediction is attributed to pressure losses, which are neglected in the inviscid analytical model but considered in the computational model. These losses increase with the height of the stack, resulting in larger discrepancies between the computational and analytical models. %

Next, the computational model is validated against a large-scale experimental study conducted by Zhang et al.~\cite{Zhang2006}. 
The experiment features a chimney with a fire source near the base, as shown in Fig.~\ref{fig:validation}(d). The chimney has a height of 2.36 m and the outside temperature is fixed at 20$^o$C. In the experiment, time variation of the temperature is measured and recorded using thermocouples placed at different heights inside the chimney.  The results are shown in Fig.~\ref{fig:validation}(e) for a thermocouple placed at a height of 2m.

The transient 3D model described in Section~\ref{sec:fluid_cfd_methodology} is adapted to mimic the conditions of the experiment whose details can be found in Ref. ~\cite{Zhang2006}. %
A comparison between the thermal responses obtained using the computational model and in the experiment is shown in Fig.~\ref{fig:validation}(e). The transient thermal response appears to be faster in the computational model than in the experiment, potentially due to uncertainty in the horizontal positioning of the thermocouples inside the chimney. However, as the solution reaches steady state, the numerical and experimental results agree within a margin of 12\%. 
Observed deviations may be due to factors such as smoke particle accumulation, wind effects at the inlet, and once again the uncertainty in thermocouple placement. 
Since engineering applications are typically concerned with the steady-state ventilation rate (rather than the short transient period, which is under one minute), this study focuses primarily on the steady-state regime, which shows good agreement with experimental observations.

\section{Results}\label{sec:results}
In this section, we analyze the influence of the origami extension on the flow field in both its expanded and contracted states. 
This is followed by an assessment of the ability of the proposed concept to achieve the key objectives of the study: (i) passive ventilation enhancement, and (ii) passive ventilation management.

\subsection{Flow-field analysis}~\label{sec:results_flow}
\begin{figure}[h!]
\centering
\includegraphics[width=\textwidth]{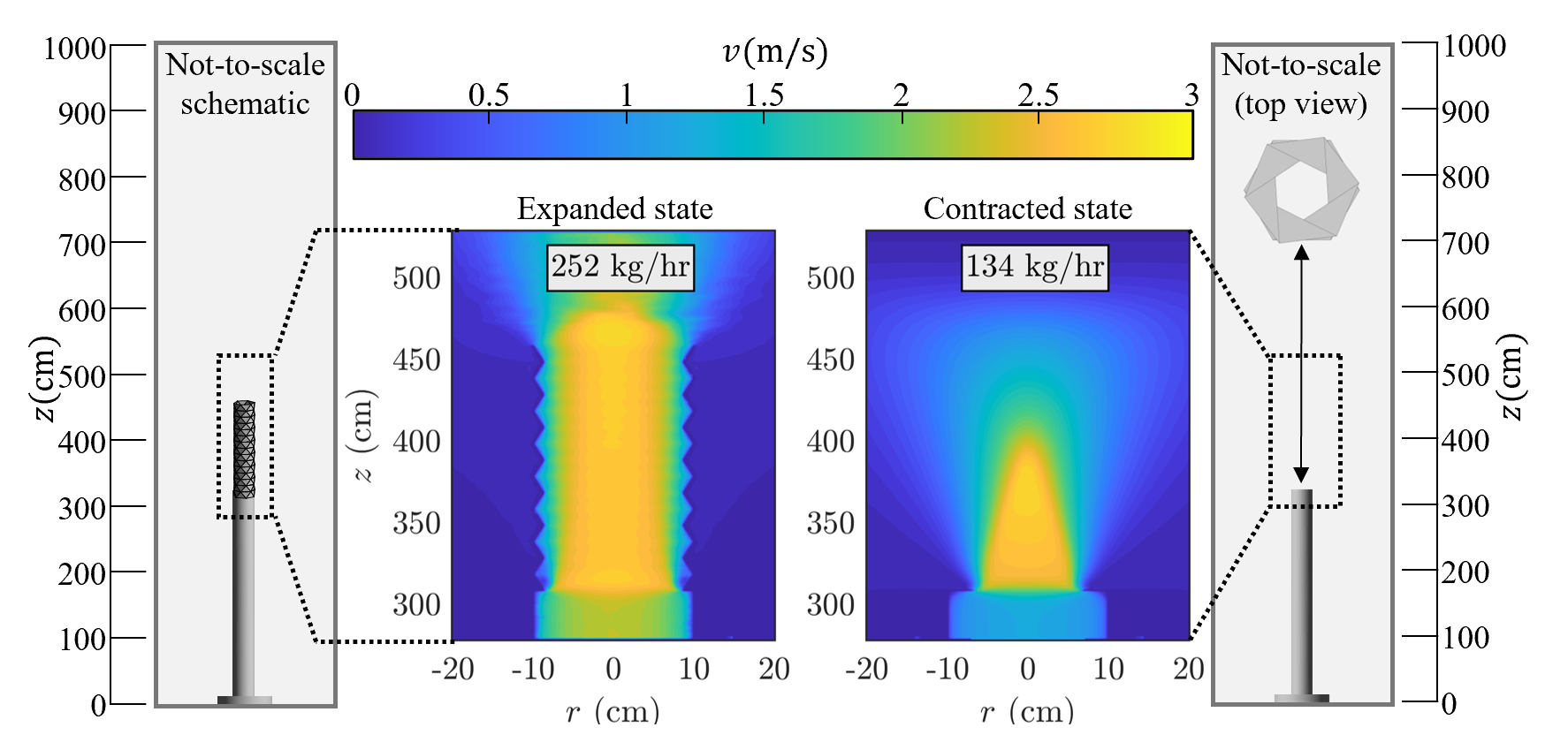}
\caption{Illustration of velocity contour plots in both expanded and contracted states.}\label{fig:illustration_contour}
\end{figure}

The velocity flow field is highly influenced by the configuration of the origami structure, whether it is in its expanded or contracted state. As illustrated in Fig.~\ref{fig:methodology}(a) and further detailed in Fig.~\ref{fig:illustration_contour}, the computational domain includes an air intake located at the bottom, followed by a heated cylindrical riser that extends up to 310 cm. An additional origami extension can increase the total height to 460 cm when fully expanded. In the contracted configuration, the origami is assumed to collapse completely to zero height (neglecting panel thickness) to simplify the geometry. This idealization is intended to evaluate the ventilation performance under optimal conditions. %

In the expanded state, the airflow exits the cylindrical riser and enters the origami structure, where it follows a zigzag trajectory due to the internal geometry of the origami. %
In the case illustrated in Fig.~\ref{fig:illustration_contour}, the structure consists of 15 stacked origami cells ($N=15$), each contributing a distinct curved boundary line, resulting in a total of 15 curved interfaces within the origami extension. %
The cross-sectional flow area within the origami is smaller than that of the riser, resulting in an increase in air velocity within the origami structure. %

In the contracted state, the origami collapses into a compact configuration that acts as a nozzle, restricting the exhaust flow before it disperses into the ambient environment. %
A comparison of air velocities at the end of the riser in both states reveals that airflow is faster in the expanded configuration than in the contracted one, indicating a higher flow rate. %
This difference is attributed to the increase in the effective stack height in the expanded state and the reduced exhaust area in the contracted state. %

Different flow patterns are observed when the design parameters $n$ and $\alpha$ of the origami unit cell are varied, as shown in Fig.~\ref{fig:contour_expanded_contracted}. These patterns can be summarized as follows:

\begin{figure}[h!]
\centering
\includegraphics[width=\textwidth]{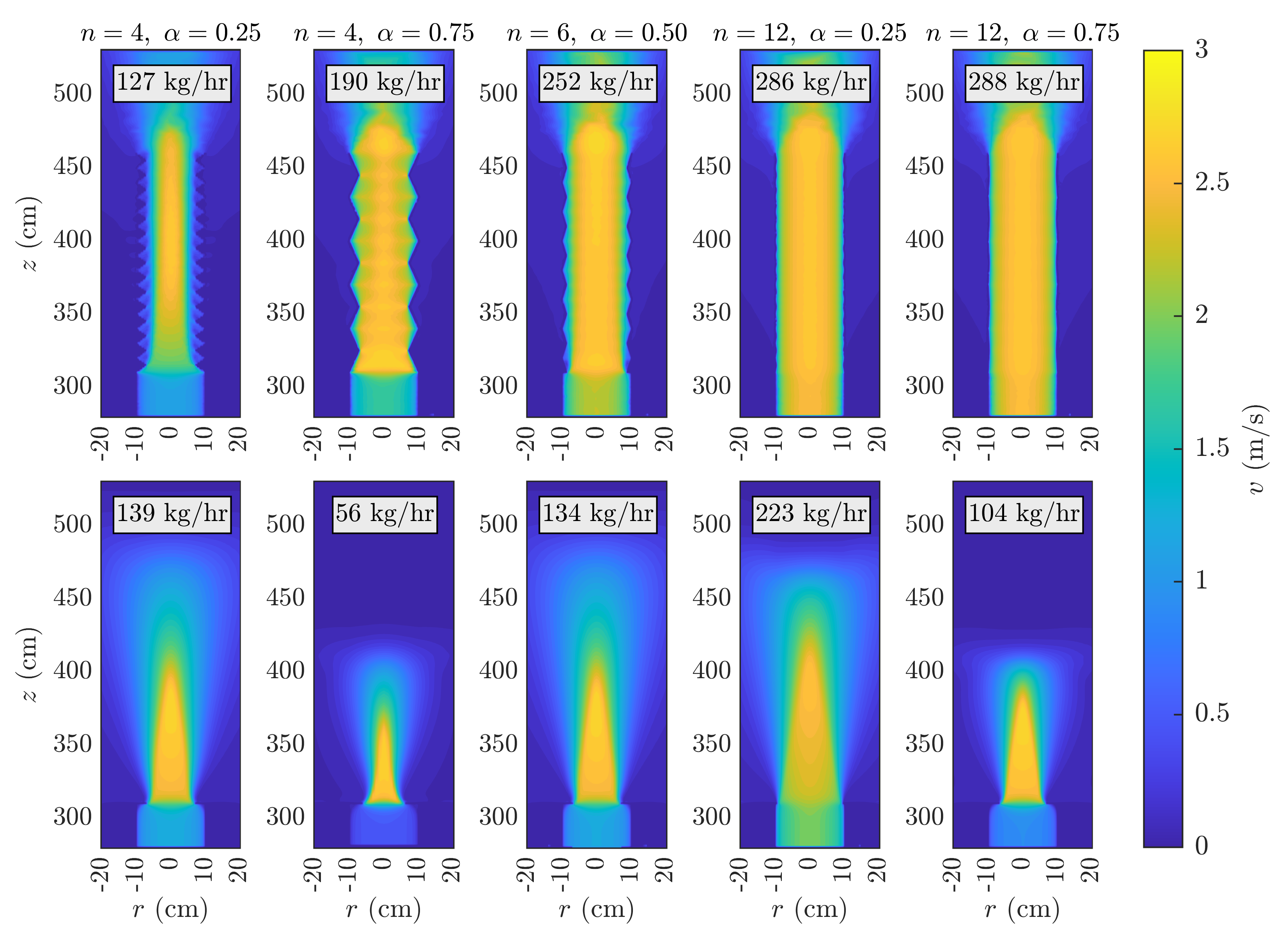}
\caption{Velocity contour plots of selected cases, with each column defined by the parameters noted at the top. 
The upper row corresponds to the expanded state, while the lower row shows the respective contracted state.
}\label{fig:contour_expanded_contracted}
\end{figure}

\begin{itemize}
\item In the expanded state, and for large values of $n$, the flow field and ventilation rate are largely unaffected by the origami aspect ratio, $\alpha$. This is because when $n$ is high, the cross-sectional shape closely approximates that of a smooth cylinder, regardless of variations in $\alpha$. As a result, the velocity contours within the origami stack and at the junction between the cylindrical riser and the origami extension are nearly identical for $\alpha = 0.25$ and $\alpha = 0.75$ when $n = 12$.

\item In the expanded state and for low values of $n$, the flow field and ventilation rate become sensitive to variations in the origami aspect ratio, $\alpha$. In this regime, the cross-section of the structure forms a polygon with $n$ sides (see Fig.~\ref{fig:appendix_origami_Cases}), and its area decreases as $n$ decreases. This reduction in cross-sectional area limits the ventilation rate, as evident by the lower airflow velocities observed in the riser region for $n=4$ and $n=6$. Additionally, a smaller value of $\alpha$ corresponds to a greater number of unit cells stacked within the same height, introducing more geometric curvature throughout the structure. This increased complexity leads to higher pressure losses, further reducing the airflow. These effects are apparent when comparing the velocity profiles in the riser for the cases of $n=4$, $\alpha=0.25$ and $n=4$, $\alpha=0.75$.

\item In the contracted state, the ventilation rate is primarily governed by the exhaust flow area. This area is proportional to $n$ and takes its maximum value (equal to that of the cylinder) when $n\rightarrow\infty$. %
For this reason, for a given $\alpha$, the ventilation rate is higher at higher values of $n$. When $n$ is fixed while $\alpha$ is increased, the flow area decreases due to the extended rotation range (see Fig.~\ref{fig:appendix_origami_Cases}).  As such, for a given $n$, the ventilation rate decreases as $\alpha$ increases, as demonstrated by the results of $n=4, \alpha=0.25$ and $n=4, \alpha=0.75$.   
\end{itemize}

Our conclusions on the impact of origami design parameters can be further observed for other combinations of $n$ and $\alpha$ in Fig.~\ref{fig:contour_expanded_appendix} and Fig.~\ref{fig:contour_contracted_appendix}.

\subsection{Ventilation enhancement}~\label{sec:results_enhancment}
The primary objective of this study is to enhance the ventilation rate by incorporating an origami extension. To evaluate the effectiveness of this approach, the mass flow rate, $\dot{m}_\textrm{e}$, through the retrofitted system with a total stack height of 4.6\,m is compared to the mass flow rate, $\dot{m}_\textrm{o}$, in the original chimney configuration with a stack height of 3\,m. 

\begin{figure}[h!]
\centering
\includegraphics{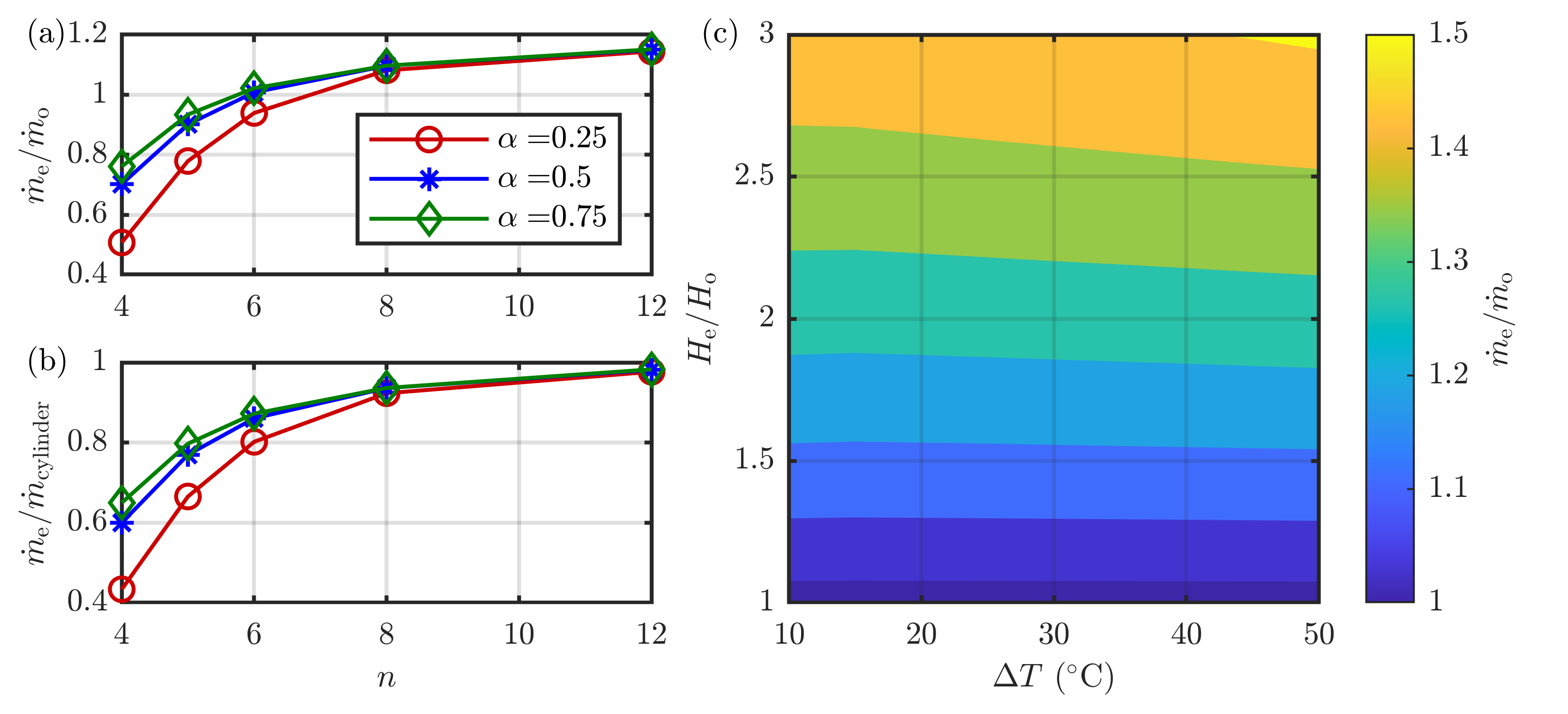}
\caption{(a) Ventilation enhancement via expanded origami cells to $H = 4.5\,\text{m}$ ($\dot{m}_\text{e}$)  with respect to baseline flow rate ($\dot{m}_\text{o}$) at original vent height of $H_\text{o} = 3\,\text{m}$ without extension, (b) ventilation rate with the same origami extension ($\dot{m}_\text{e}$) but compared to a ventilation rate of a smooth cylinder, ($\dot{m}_\text{cylinder}$) reaching the same stack height of $H=4.5\text{m}$, and (c) overall ventilation enhancement expressed as a function indoor-outdoor temperature difference, $\Delta T$, and non-dimensional height, $H_\text{e}/H_\text{o}$.}\label{fig:enhancement}
\end{figure}

It is evident from Fig.~\ref{fig:enhancement}(a) that both $n$ and $\alpha$ have a significant impact on the ventilation rate. For lower values of $n$ (up to $n = 5$), ventilation performance declines despite the increased stack height. This reduction is primarily due to two factors: $i)$ a smaller exhaust flow area compared to the original smooth cylindrical riser, and $ii)$ increased flow resistance caused by the highly-curved panel geometry. However, as $n$ increases beyond 6, the mass flow rate improves considerably and surpasses $\dot{m}_\text{o}$. 

The effect of the cell aspect ratio, $\alpha$, becomes critical at low values of $n$ due to increased pressure losses, as previously discussed in Section~\ref{sec:results_flow} and illustrated in Fig.~\ref{fig:contour_expanded_contracted}.  
To better assess the system's performance, the flow rate through the origami extension,  $\dot{m}_\textrm{e}$, is also compared to the flow rate, $\dot{m}_\textrm{cylinder}$,  through a smooth cylindrical stack of equal height (4.6\,m), as shown in Fig.~\ref{fig:enhancement}(b).  
A system efficiency greater than 90\% can be achieved with just eight panels. Increasing $n$ can further improve system efficiency and reduce sensitivity to the cell aspect ratio, albeit at the cost of increased mechanical complexity.  

To investigate the impact of stack height on ventilation rate under varying environmental conditions, the flow rate ratio $\dot{m}_\textrm{e}/\dot{m}_\text{o}$, is calculated for stack heights, $H$, ranging from 3m to 9m (corresponding to $H^* = H_\text{e}/H_\text{o} = 1 \rightarrow 3$) and indoor-outdoor temperature differential ranging from 10$^\circ$C to 50$^\circ$C as shown in Fig.~\ref{fig:enhancement}(b). %
The results demonstrate that retrofitting an origami extension on top of an existing riser can lead to a substantial enhancement in ventilation performance upon expansion. 
In particular, approximately 25\% enhancement in the flow rate ratio is observed every time the original height is doubled. The improvement is more pronounced at higher values of $\Delta T$, where the buoyancy forces are larger and help accelerate the airflow through the origami stack. %

\subsection{Ventilation management}~\label{sec:results_management}
\begin{figure}[h!]
\centering
\includegraphics{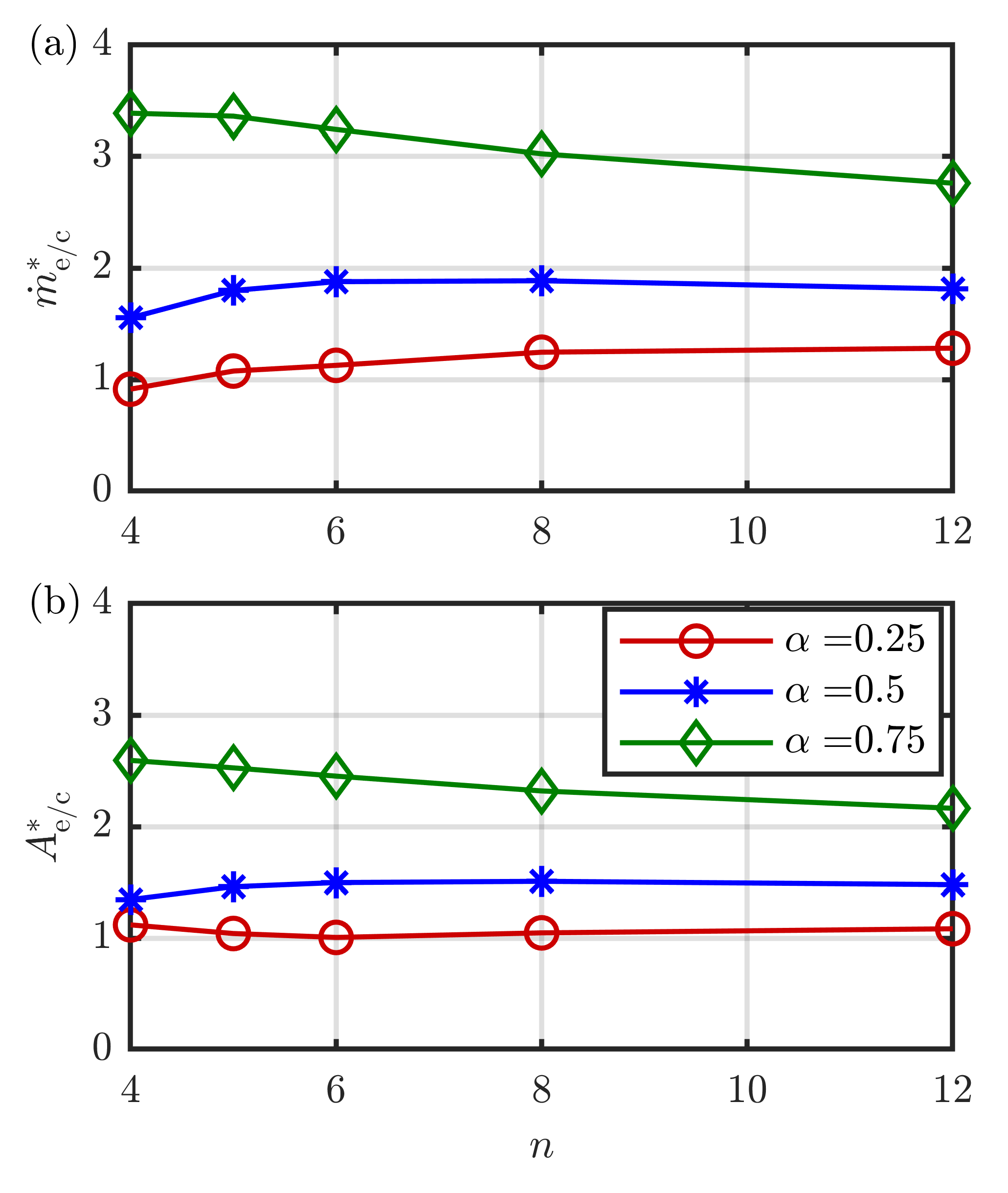}
\caption{(a) Ventilation management expressed as $\dot{m}_\text{e/c}^*=\dot{m}_\textrm{expanded}/\dot{m}_\textrm{contracted}$, and (b) the ratio of flow area in expanded and contracted states, $A_{\text{e/c}}^*=A_\textrm{expanded}/A_\textrm{contracted}$.}\label{fig:tunability}
\end{figure}

Another objectives of this study is to demonstrate how origami structures can be used to control ventilation rates on demand through controlled expansion and contraction, without the need for additional mechanical components. To this end, we introduce the ratio, $\dot{m}_\text{e/c}$, which measures the ventilation rate in the expanded state divided by that in the contracted state. As shown in Fig.~\ref{fig:tunability}, the ratio $\dot{m}_\text{e/c}$ was evaluated for 30 combinations of $n$ and $\alpha$.  Results shown in Fig. ~\ref{fig:tunability}(a) indicate that $\dot{m}_\text{e/c}$ is much more sensitive to variation in the cell aspect ratio, $\alpha$, than it is to variations in $n$. In particular, higher values of $\alpha$ enable greater controllability of the ventilation rate $\dot{m}_\text{e/c}$. 
This can be attributed to the larger allowable cell rotation; i.e. rotation of the upper polygon as the unit cell contracts from the fully expanded to the fully contracted position.  The larger this rotation, the more significant the changes in flow area during expansion and contraction, as calculated in Fig.~\ref{fig:tunability}(b) and visualized in Fig.~\ref{fig:appendix_origami_Cases}. %
Quite interestingly, using a cell aspect ratio of $\alpha=0.75$ can control ventilation rates by approximately a factor of three.
It is important to note that this ratio can theoretically increase to infinity if $\alpha$ is further increased such that the contracted flow area approaches zero. %

The number of triangular panels ($2n$) has a secondary influence on $\dot{m}_\text{e/c}$. In general, changing the number of panels results in a change $\dot{m}_\text{e/c}$ that is proportional to the change in flow area.
In the special case of low $n$ and $\alpha$, the increase in pressure losses of the expanded case becomes significant resulting in a reduction in the value of $\dot{m}_\text{e/c}$. 
However, the aspect ratio remains generally the dominant parameter in achieving passive ventilation control. 
Consequently, origami ventilation designs with high values of $\alpha$ can independently control the ventilation rate, while those with low values of $\alpha$ may require additional mechanisms to regulate airflow effectively.

\section{Conclusion}\label{sec:conclusion}
This study introduces a deployable origami-based system designed to enhance and regulate passive ventilation through the stack effect. The approach involves retrofitting a chimney or riser with a Kresling origami extension that passively adjusts both the stack height and the exhaust flow area, the two key parameters governing buoyancy-driven airflow.

To evaluate the system’s performance, both computational and analytical models are developed, examining the influence of critical design variables such as the origami cell aspect ratio, the number of triangular panels used to construct the unit cell, and the overall stack height. The key findings are summarized as follows:

\begin{itemize}
    \item Around threefold ventilation tunability is demonstrated, with potential to increase tunability further as the cell aspect ratio is increased. 
    \item Around 25\% enhancement in ventilation rate with every doubling of stack height.
    \item Origami cells with a minimum of six panels are required to achieve efficient ventilation enhancement, avoiding excessive pressure drop and reduced exhaust flow areas.
\end{itemize}

The findings suggest that origami-based systems offer a promising pathway for adaptive ventilation management in extreme and off-grid environments. Future work will focus on developing an experimental setup to validate the computational findings.

\appendix
\section{Computational Model Formulation}\label{Appendix:computational}
The airflow through the origami-based ventilation system was modeled using a three-dimensional, transient formulation of the Reynolds-Averaged Navier–Stokes (RANS) equations, with air treated as an ideal gas. The governing conservation equations are:\\

\noindent \underline{\textbf{Mass Conservation:}}
\begin{equation}
\frac{\partial \rho}{\partial t} + \nabla \cdot (\rho \mathbf{u}) = 0
\end{equation}
where \( \rho \) is the air density [kg/m\textsuperscript{3}], \( t \) is time [s], and \( \mathbf{u} \) is the velocity vector [m/s].

\noindent \underline{\textbf{Momentum Conservation:}}
\begin{equation}
\frac{\partial (\rho \mathbf{u})}{\partial t} + \nabla \cdot (\rho \mathbf{u} \mathbf{u}) = -\nabla P + \nabla \cdot \left[\mu_{\text{eff}} \left( \nabla \mathbf{u} + (\nabla \mathbf{u})^T \right) \right] + \rho \mathbf{g}
\end{equation}
where \( P \) is pressure [Pa], \( \mu_{\text{eff}} \) is the effective viscosity [Pa·s], including both molecular and turbulent contributions, and \( \mathbf{g} \) is the gravitational acceleration vector [m/s\textsuperscript{2}].

\noindent \underline{\textbf{Energy Conservation:}}
\begin{equation}
\frac{\partial (\rho h)}{\partial t} + \nabla \cdot (\rho \mathbf{u} h) = \nabla \cdot (\lambda_{\text{eff}} \nabla T) + \Phi
\end{equation}
where \( h \) is the specific enthalpy [J/kg], \( T \) is the temperature [K], \( \lambda_{\text{eff}} \) is the effective thermal conductivity [W/m·K], which includes both molecular and turbulent contributions, and \( \Phi \) is an energy source term included in the validation case against the experimental field data of~\cite{Zhang2006}, which involved a fire source.

The ideal gas law is used to couple density, pressure, and temperature:
\begin{equation}
\rho = \frac{P}{R T}
\end{equation}
where \( R \) is the specific gas constant for air [J/kg·K].

The Shear Stress Transport (SST) \(k\)-\(\omega\) model was used to capture turbulence effects due to the high Reynolds number of the flow (\(Re > 10^5\)). The transport equations for turbulent kinetic energy \(k\) and specific dissipation rate \(\omega\) are:

\noindent \underline{\textbf{Turbulent Kinetic Energy:}}
\begin{equation}
\frac{\partial (\rho k)}{\partial t} + \nabla \cdot (\rho \mathbf{u} k) = \nabla \cdot \left[ \left( \mu + \frac{\mu_t}{\sigma_k} \right) \nabla k \right] + P_k - \beta^* \rho k \omega
\end{equation}
where \( k \) is the turbulent kinetic energy [m\textsuperscript{2}/s\textsuperscript{2}], \( \mu \) is the dynamic viscosity [Pa·s], \( \mu_t \) is the turbulent eddy viscosity [Pa·s], \( \sigma_k \) is a model constant, \( P_k \) is the production of turbulent kinetic energy [W/kg], and \( \beta^* \) is a model constant.

\noindent \underline{\textbf{Specific Dissipation Rate:}}
\begin{equation}
\begin{aligned}
\frac{\partial (\rho \omega)}{\partial t} + \nabla \cdot (\rho \mathbf{u} \omega) &= \nabla \cdot \left[ \left( \mu + \frac{\mu_t}{\sigma_\omega} \right) \nabla \omega \right] + \alpha \frac{\omega}{k} P_k \\
&\quad - \beta \rho \omega^2 + 2(1 - F_1) \rho \sigma_{\omega 2} \frac{1}{\omega} \nabla k \cdot \nabla \omega
\end{aligned}
\end{equation}
where \( \omega \) is the specific dissipation rate [1/s], \( \sigma_\omega \), \( \alpha \), \( \beta \), and \( \sigma_{\omega 2} \) are model constants, and \( F_1 \) is a blending function.

The eddy viscosity \( \mu_t \) is computed as:
\begin{equation}
\mu_t = \frac{\rho k}{\omega}
\end{equation}
\noindent \underline{\textbf{Boundary Conditions:}}
\begin{itemize}
    \item \textbf{Walls:} No-slip velocity condition (\( \mathbf{u} = 0 \)); adiabatic thermal condition (\( \partial T/\partial n = 0 \))
\end{itemize}

\noindent \underline{\textbf{Initial Conditions:}}
\begin{itemize}
    \item Prescribed indoor and outdoor temperatures depending on the simulation case
    \item Pressure: \( p = 1~\text{atm} \)
    \item Velocity: \( \mathbf{u} = 0 \)
\end{itemize}
Note that, except for the validation model against field data of Zhang et al.~\cite{Zhang2006} in Section~\ref{sec:methodology_val}, the indoor temperature is fixed at a typical value of 20$^\circ$C to drive the buoyant ventilation.

\section{Results: Velocity contours}
\setcounter{figure}{0}
\newpage
\subsection{Velocity contours in expanded state}
\begin{figure}[h!]
\centering
\includegraphics[width=12.5cm]{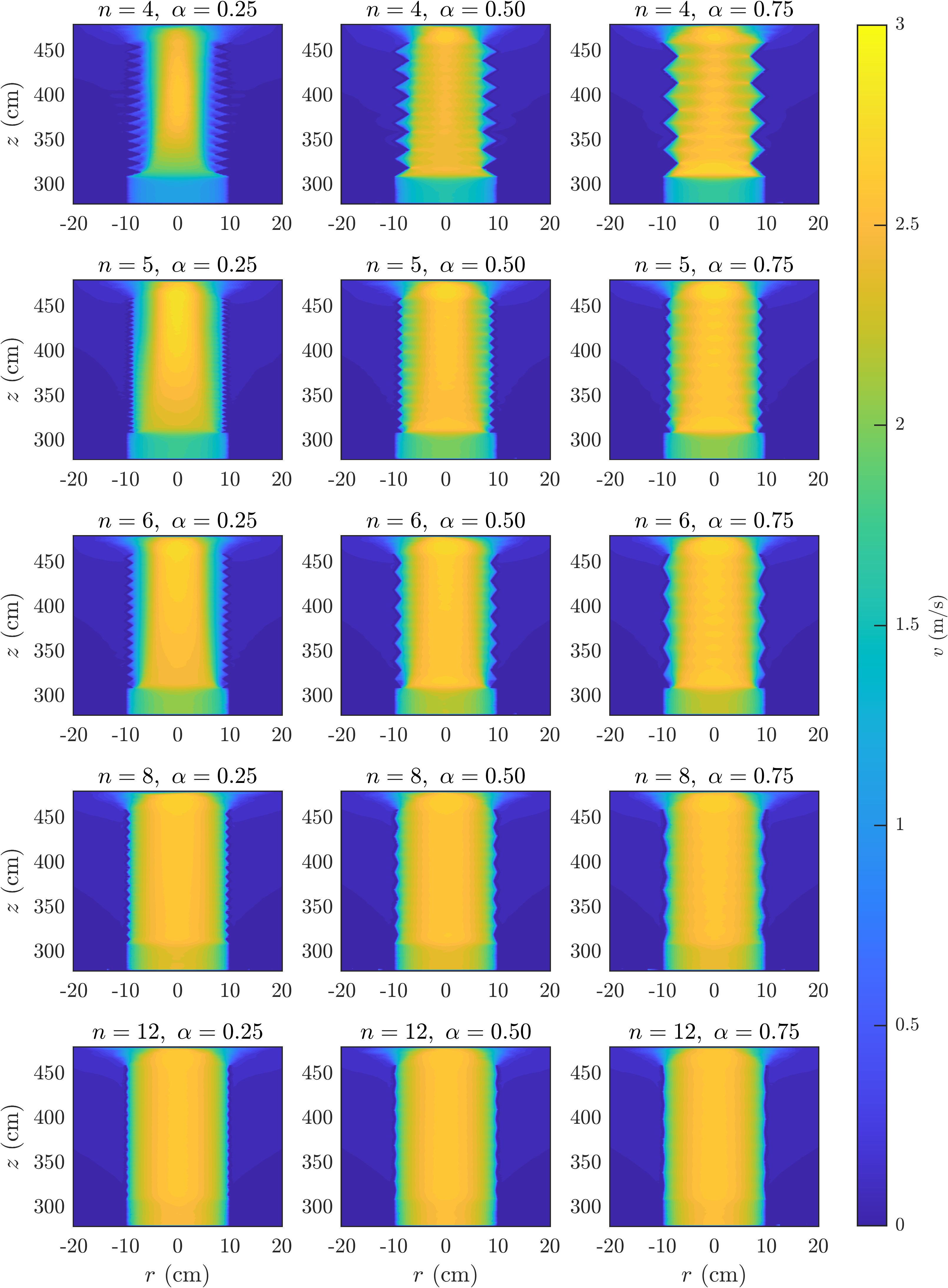}
\caption{Velocity contours of all cases in expanded state.}\label{fig:contour_expanded_appendix}
\end{figure}
\newpage
\subsection{Velocity contours in contracted state}
\begin{figure}[h!]
\centering
\includegraphics[width=12.5cm]{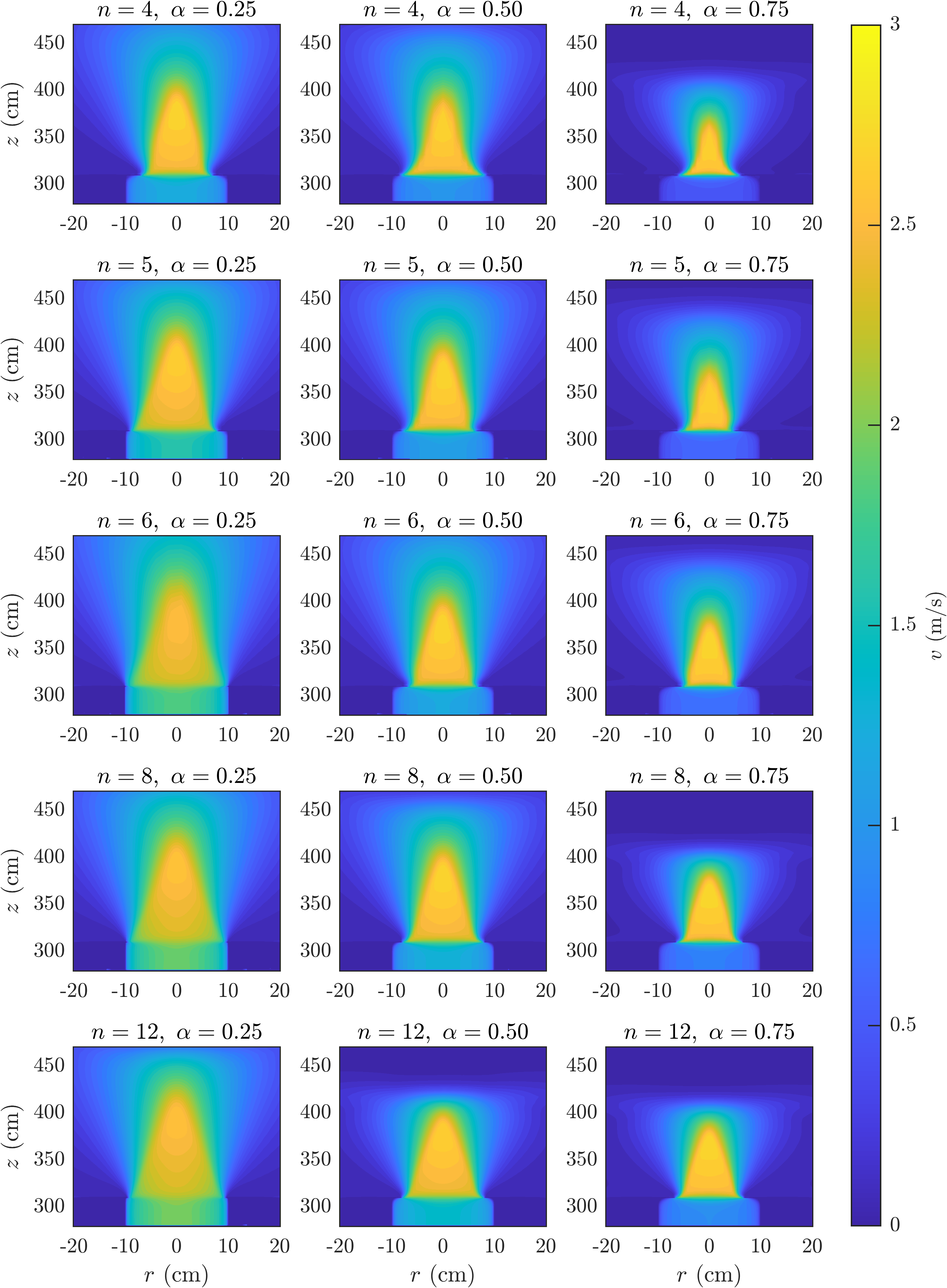}
\caption{Velocity contours of all cases in contracted state.}\label{fig:contour_contracted_appendix}
\end{figure}

\section*{Acknowledgments}
This research was supported by New York University IT’s High Performance Computing resources, services, and expert staff.
\nocite{*}

\bibliographystyle{elsarticle-num}
\bibliography{export.bib}

\end{document}